\documentclass[aps,prl,twocolumn,superscriptaddress,showpacs]{revtex4-1}
\usepackage{graphicx}
 \usepackage{psfrag}
\usepackage{ae}
\usepackage{amsmath,amssymb}
\usepackage{float}

\begin{document}
\date{\today}

\title{Stability and self-organization of planetary systems}

\author{Renato Pakter}
\author{Yan Levin}
\affiliation{
  Instituto de F\'{\i}sica, UFRGS, 
  Caixa Postal 15051, CEP 91501-970, Porto Alegre, RS, Brazil  }


%

\begin{abstract}
We show that stability of planetary systems is intimately connected with their internal order.
An arbitrary initial distribution of planets is susceptible to catastrophic events in which
planets either collide or are ejected from the planetary system.  These instabilities are
a fundamental consequence of chaotic dynamics and of Arnold diffusion characteristic of many body  gravitational interactions.  To ensure stability over astronomical time scale of a {\it realistic} planetary system  -- in which planets have masses comparable to those of planets in the solar system -- the motion must be quasi-periodic. 
A dynamical mechanism is proposed which naturally evolves a planetary system to a quasi-periodic state from an arbitrary initial condition. A planetary 
self-organization predicted by the theory is similar to the one found in our solar system.  
\end{abstract}

\maketitle

\section{Introduction}
\begin{figure}[h]
\begin{center}
\includegraphics[scale=0.325]{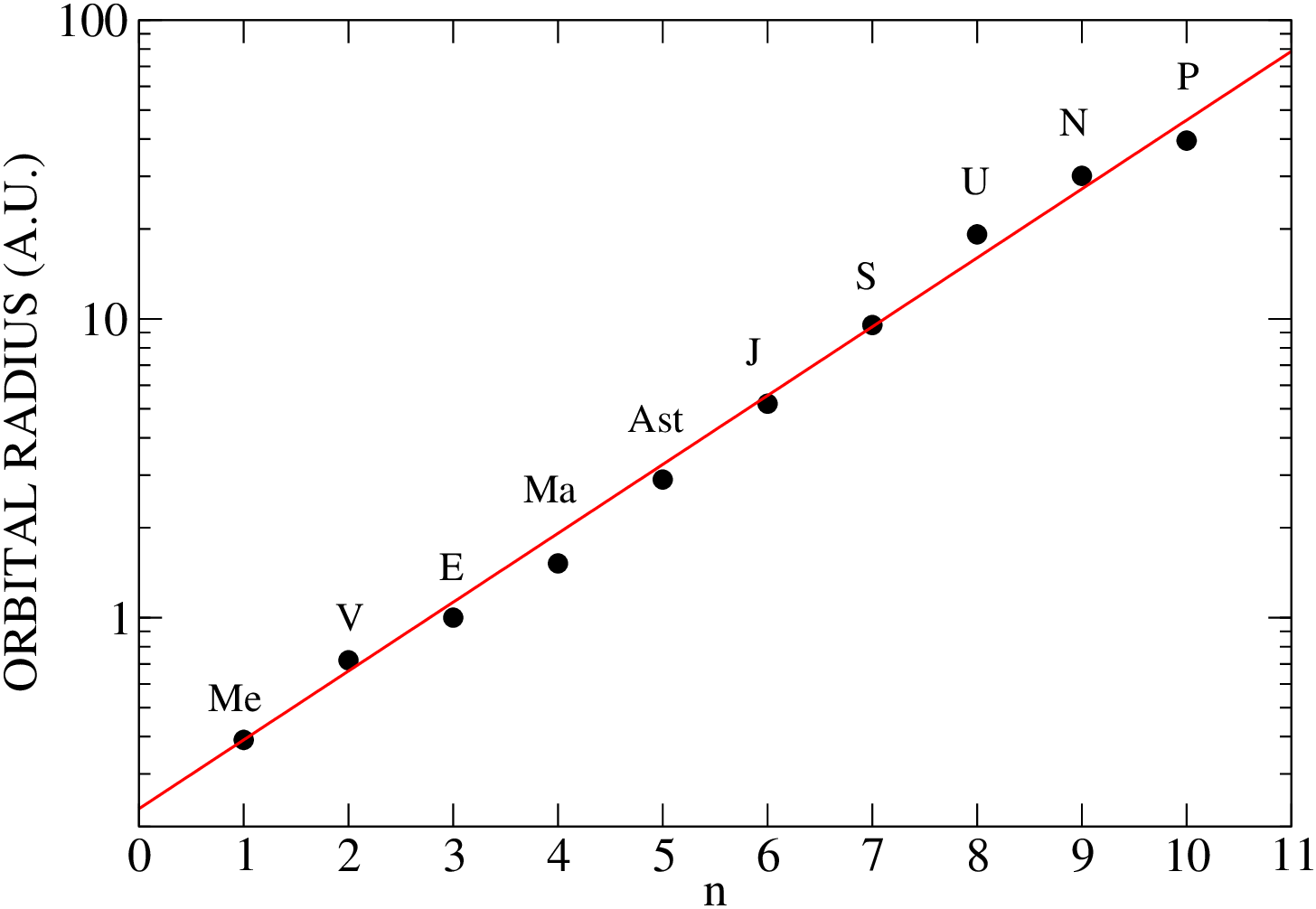}
\caption{Mean radii of planets in our solar system on a semi-log scale: (Me) Mercury, (V) Venus, (E) Earth, (Ma) Mars, (Ast) Asteroid belt, (J) Jupiter, (S) Saturn, (U) Uranus, (N) Neptune, (P) Pluto.  Straight line is the modern version of Titius-Bode law.} \label{fig1}
\end{center}
\end{figure}

Stability of solar system has been a subject of great debate 
since the time that Isaac Newton first wrote his famous law of universal gravitation~\cite{La13}.
Newton realized that the interaction between planets will perturb 
their orbits from a simple elliptical shape postulated by Kepler.  Over millions of years
this small perturbations will accumulate leading to catastrophic events such as collisions between planets or ejection of planets from the solar system \cite{New}.  Newton's solution to this difficulty was to invoke Divine Intervention in which God would  adjust the planetary orbits to keep the solar system stable.  Newton's contemporary, Leibnitz, 
objected strongly to Newton's theological solution: ``Was God an inferior watchmaker, he demanded, who could not get things right from the beginning?" \cite{Mor} Over the centuries the question of stability has attracted attention of both physicists and mathematicians, without a definite solution.  
An apparently unrelated  
puzzle is a seeming regularity of spacing of planetary orbits~\citep{HaTr98}, see Fig. \ref{fig1}. In 1766 Titius
noticed that the orbits of then known planets followed a geometric sequence,
if a ``missing" planet was inserted between Mars and Jupiter~\citep{Ni72}. Subsequent discovery of the asteroid belt and of Uranus at the positions predicted by the Titius-Bode ``law" gave further credence to the belief that there is a hidden order in the solar system. The modern version of the Titius-Bode law can be written as $r_n=r_0 1.7^n$, with $n=1$ being Mercury, $n=2$ Venus, $n=3$ Earth, etc., see Fig. \ref{fig1}. 
The normalization $r_0=0.2294$ A.U. was chosen so that $n=1$ corresponds exactly to the orbit of Mercury in Astronomical Units (A.U.) \cite{GrDu94}. 

In this paper we will show that stability of planetary systems is intimately connected with their internal order.  The dynamical simulations demonstrate that a generic arrangement of planets is unstable to small perturbations resulting from interplanetary interactions which lead to catastrophic events.  We argue that a planetary system will remain stable over astronomical time scales only if its dynamics is quasi-periodic. 
Indeed various near-commensurabilities have been observed for satellites and planets in the solar system~\cite{Go64,Pe76} and already in 1970 Hill suggested that there is a dynamical origin to the Titius-Bode law~\cite{Hi70}. 
In this paper we will argue that a specific requirement of quasi-periodicity
results in a planetary distribution almost identical to the one observed in the solar system.  Furthermore, we 
will provide a dynamical mechanism that leads to spontaneous self-organization of a planetary system into a periodic state.  This is different from other approaches used previously to explain mass distribution in the solar system which rely either on statistics or hydrodynamic instabilities~\cite{GrDu94,La00,HeBe11,LaPe17}.  Even if such approaches can account for geometrical progression of planetary distances, they can not explain stability of planetary systems.  
The Titius-Bode law is not a condition of stability, but rather a consequence of self-organization, as will be demonstrated in the present paper. Finally we should mention that according to Kolmogorov-Arnold-Moser (KAM) theorem there is a dense set of initial conditions for a many body 
gravitational system that lead to a stable quasi-periodic trajectories. However, as was already demonstrated by Henon in 1966,
the KAM stability only applies to planetary systems with unrealistically small planetary masses of less than 
$10^{-320}$ of solar mass~\cite{La14}.  A generic initial condition for a realistic planetary system will, therefore, result in a chaotic dynamics, such that  in the infinite time limit planets will either collide or will be ejected from the system.

\section{Theory}

\begin{figure}
\begin{center}
\includegraphics[scale=0.325]{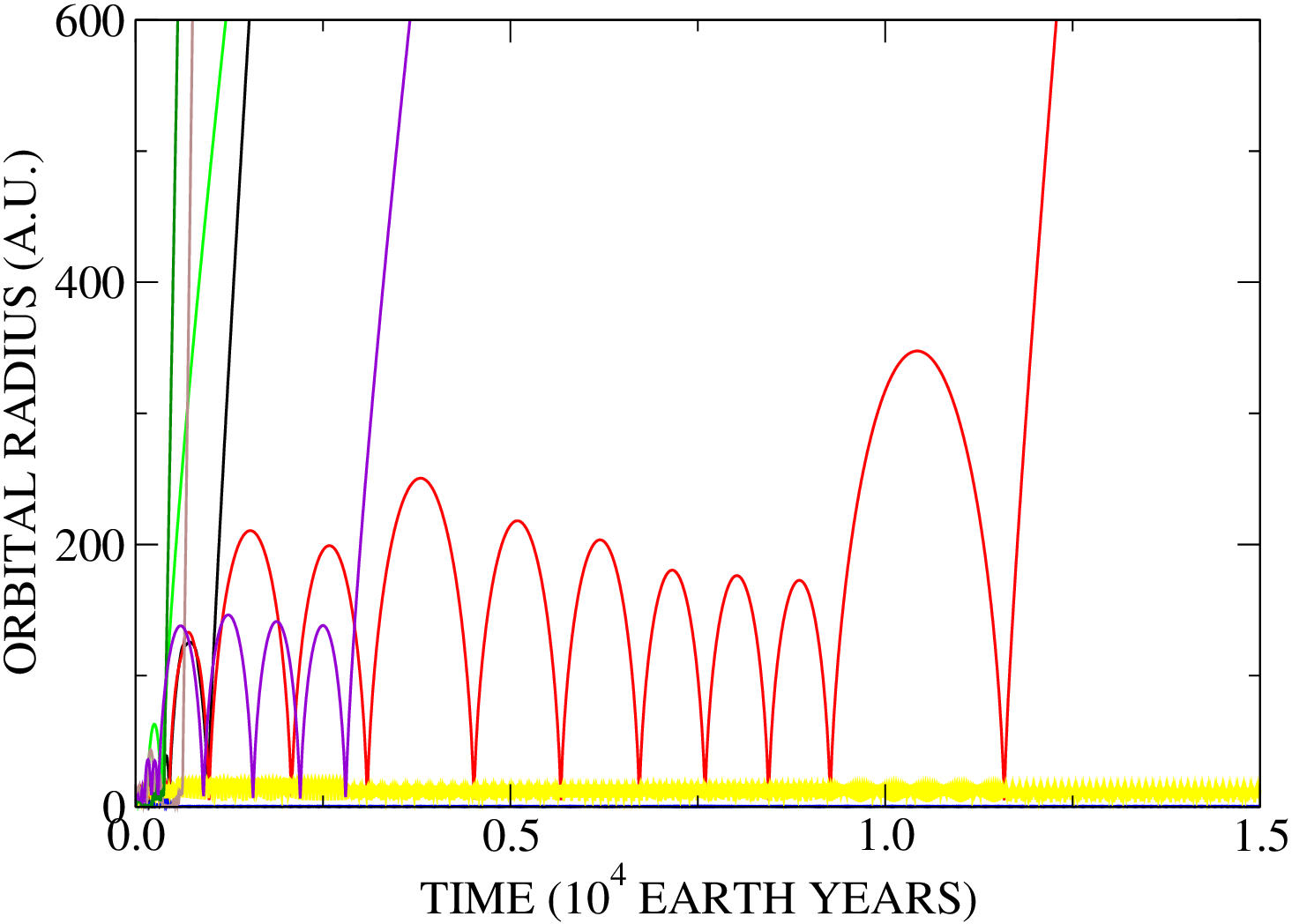}
\caption{Radial coordinate evolution of $8$ planets of mass $0.005M$ originally distributed uniformly between $1$ and $8$ A.U. in their respective Keplerian orbits.  After a short time we start seeing catastrophic events (the almost vertical trajectories) in which planets begin to be ejected from the planetary system.  The resulting planetary system remains with only $2$ planets.} \label{fig2}
\end{center}
\end{figure}

The stability of planetary systems is an outstanding problem.  
Since Newton's gravitational potential is bound
from above and is unbound from below \cite{LePa14,BeRi14}, orbits of 
planets are in general unstable --  some planets can 
gain enough kinetic energy to escape altogether from the planetary system~\cite{ChWe96}, 
while others will 
fall into sun or collide with each other, see Fig. \ref{fig2}.  This type of instability driven by chaos and Arnold diffusion \cite{Ar64}, is a fundamental characteristic of many body celestial dynamics. On the other hand it is possible to find very special initial conditions  --- correspond to a set of measure zero, since KAM theorem  does not apply to realistic planetary systems \cite{La14} ---  for which the dynamics of a non-linear interacting system is purely periodic.  This periodic solutions will persist indefinitely, with the relative configuration of planets repeating itself after a synodic period of time $T$.  The fundamental difficulty is to obtain initial conditions which lead to periodic dynamics of a fully interacting gravitational system.  
Furthermore, since the probability that a planetary system  will be  ``born" with planets at precisely the correct positions is highly improbable, there must be a mechanism that makes a planetary system evolve towards a stable periodic orbit.  Motivated by the theories of control of chaos~\cite{OtGr90,Py92}, we suggest that periodic orbits can be stabilized by energy non-conserving perturbations~\cite{JoMo16}. Such perturbations could have originated from the interaction of newborn planetesimals with the gas and dust of the protoplanetary disk.  The angular frequency of dust particles at a distance $r$ from the star has a simple Keplerian form~\cite{ShSu73,LyPr74}
\begin{eqnarray}
\label{e1}
\omega(r)= \sqrt{\frac{G M}{r^3}} ,
\end{eqnarray}
where  $M$ is the star mass and $G$ is Newton's gravitational constant. For concreteness we will take the star mass to be that of our sun and will measure all the distances in astronomical units and time in earth years. To simplify the calculations we will suppose that all the planetary orbits are restricted to the ecliptic plane and all planets have the same mass $m$.  To speed up the simulations we will take the planetary mass 
$m$ to be a few times that of Jupiter. We will see, however, that our conclusions do not depend on the planetary mass as long as it is much smaller than the star mass.  

Newton's equations of motion for the coordinates $x$ and $y$ of a planet $i$ are
\begin{eqnarray}
\label{e2}
m\ddot x_i=-\frac{GMm x_i}{r_i^3}-\sum_j\frac{Gm^2 (x_i-x_j)}{r_{ij}^3}-f^\theta_i \frac{y_i}{r_i} \\ \nonumber 
m\ddot y_i=-\frac{GMm y_i}{r_i^3}-\sum_j\frac{Gm^2 (y_i-y_j)}{r_{ij}^3}+f^\theta_i \frac{x_i}{r_i} \;,
\end{eqnarray}
where $r_i=\sqrt{x_i^2+y_i^2}$ is the distance of $i$'th planet from the star and
$r_{ij}=\sqrt{(x_i-x_j)^2+(y_i-y_j)^2}$ is the separation 
between planet $i$ and planet $j$. The purely angular force ${\pmb f}^\theta_i=f^\theta_i \hat{\pmb \theta}$, where $\hat{\pmb \theta}$ is the unit angular vector, is responsible 
for the interaction of a planet $i$ with the residual dust of the protoplanetary disk.
Here we will use a simple phenomenological expression for such non-conservative force:
if the angular velocity of a planet is lower than the Keplerian velocity of the surrounding dust in the same orbit, the planet will gain energy from dust; if the planetary velocity is larger than the velocity of dust, it will loose energy.  The simplest possible mathematical expression with this characteristic is:
\begin{equation}
\label{e3}
f^\theta_i=-\beta \left (r_i \omega(r_i)- v^\theta_i \right)\left[L-m \sum_i v^\theta_i r_i \right]  \,,
\end{equation}
where $\beta$ is a small phenomenological constant which controls the interaction between dust and planets, $v^{dust}_i=r_i \omega(r_i)$ is the angular velocity of dust at the location of planet $i$, and $v^\theta_i$ is the angular velocity of the planet $i$, 
\begin{eqnarray}
\label{e4}
v^\theta_i=\frac{\dot y_i x_i-y_i \dot x_i}{r_i} \,.
\end{eqnarray}
The expression Eq. (\ref{e3}) is analogous to viscous dissipation of an object in a rotating fluid.  
The term in square brackets of Eq.(\ref{e3}) is included so that the non-conservative force ``turns off"  when the net planetary angular momentum reaches a predetermined value $L$.  This is designed to model a progressive depletion of dust/gas from the protoplanetary disk which will result in a continuously decreasing value of $f^\theta$. In our simulations we used $\beta=10^{-3}-10^{-2}$ A.U.$^{-2}$.  To integrate the equations numerically we employed a Runge-Kutta algorithm with adaptive time-step that uses embedded fifth
order and sixth order Runge-Kutta estimates to compute the solution and the relative error~\cite{NumRecipesC}. To speed up the simulation and avoid numerical instabilities the singular form of the Newton's gravitational potential between the planets is regularized to,
\begin{eqnarray}
\label{e2a}
V(r)&=&-\frac{(2 d^3 - 2 d r^2 + r^3) G m^2}{d^4} \,\, \textrm{for} \,\,r \le d \\ \nonumber 
V(r)&=&-\frac{G m^2}{r} \,\, \textrm{for}  \,\,r > d\;,
\end{eqnarray}
where $d$ is an arbitrary short distance cutoff whose precise value does not affect our conclusions. In our simulations we used $d$ on the order of $10^{-3}$ A.U. With this modification, we do not need to introduce any specific ``collision model", since the integrator will have enough resolution to deal with the fast dynamics resulting from catastrophic planetary collisions. 
Indeed, the dynamics of planets interacting through Eq. (\ref{e2a})  can result in planetary collision in which  two or more trajectories merge into one.  We shall call such events ``planetary coalescence", which are analogous to non-elastic collisions. The advantage of regularized Newton potential Eq.(\ref{e2a}) is that such singular events can be handled using the adaptive step size Runge-Kutta integrator without any numerical instabilities. 

\begin{figure}
\begin{center}
\includegraphics[scale=0.325]{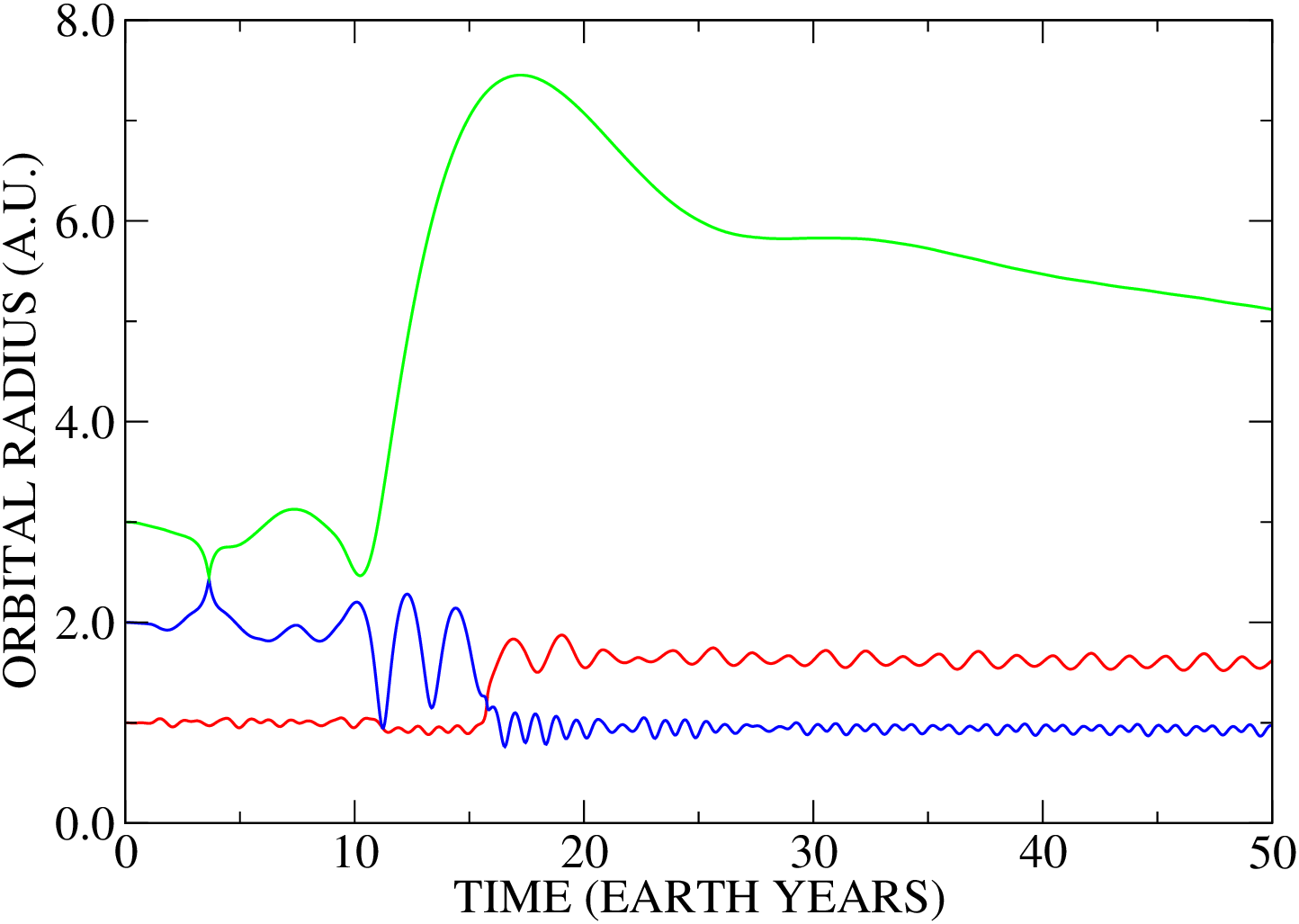}
\caption{Temporal evolution of radial coordinates of 3 planets under the action of Eqs. (\ref{e3}).  The dynamics is very complex with two of the planets switching their relative order in the sequence.
} \label{fig2a}
\vspace{0.7cm}
\end{center}
\end{figure}

We stress that our goal is to find the simplest possible mechanism that can drive a planetary system towards a stable quasi-periodic state.  During the formation of solar system, more complicated gravitational, electromagnetic, and collisional processes have certainly taken
place.  Here, however, our aim is to provide a proof of concept that energy non-conserving perturbation can drive
a planetary system into a self-organized quasi-periodic state.
In the simulations we 
observe that the angular 
force given by Eq. (\ref{e3}) leads to a self-organized periodic state if $\beta \rightarrow 0$ and $t \rightarrow \infty$.  In practice, however, the simulation time is finite,
so that $\beta$ can not be too small.  In Fig. \ref{fig2a} we show the dynamical evolution described by Eqs. (\ref{e2}) of a 
system with 3 planets, initially placed in Keplerian orbits uniformly distributed from $1$ A.U. to $3$ A.U. 

\begin{figure}
\begin{center}
\includegraphics[scale=0.325]{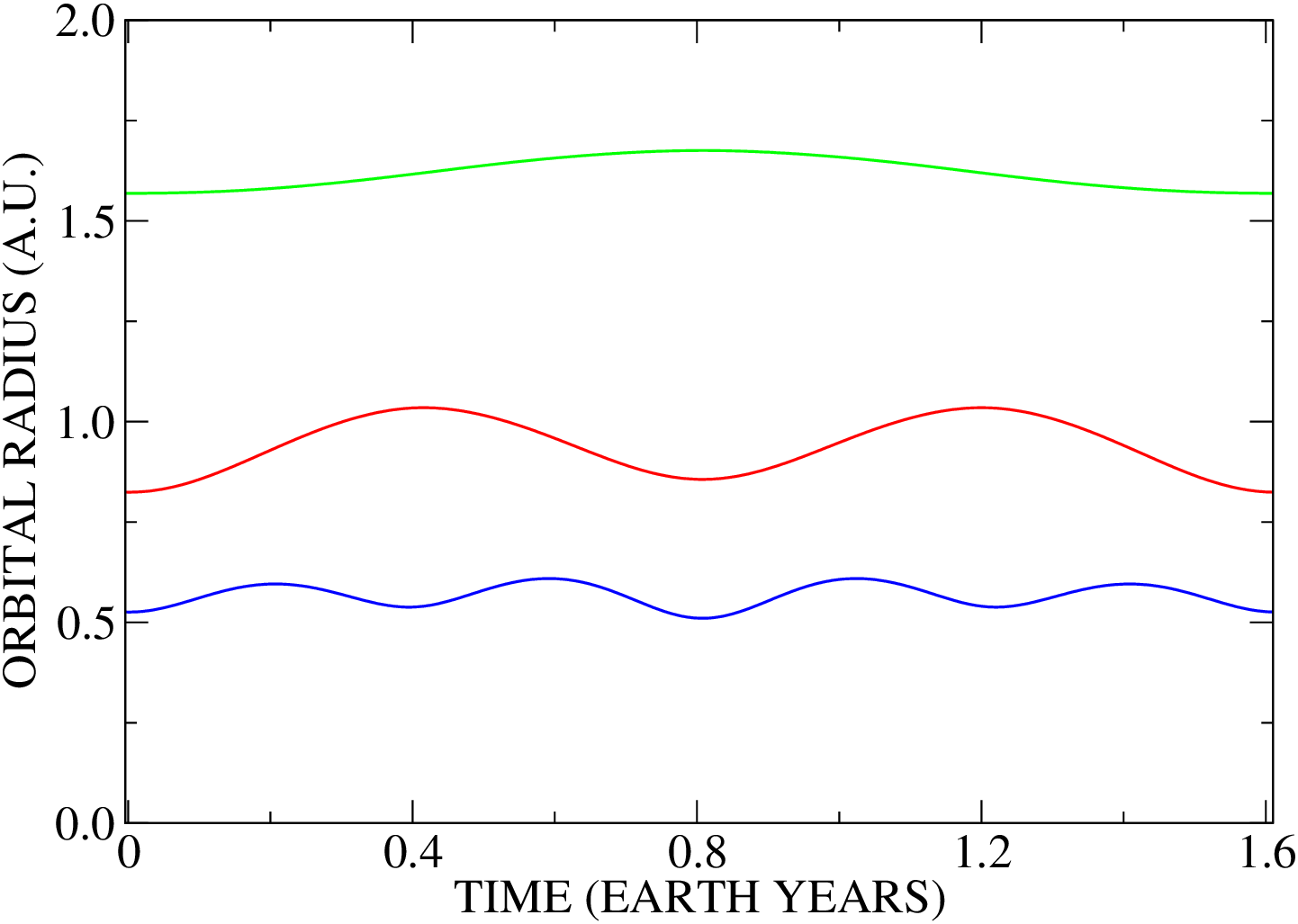}
\caption{Temporal evolution of the radial coordinates of the $3$ planet system of Fig. \ref{fig2a} 
after a self-organized 
periodic state has been established.
} \label{fig2b}
\vspace{0.7cm}
\end{center}
\end{figure}

\begin{figure}
\begin{center}
\includegraphics[scale=0.325]{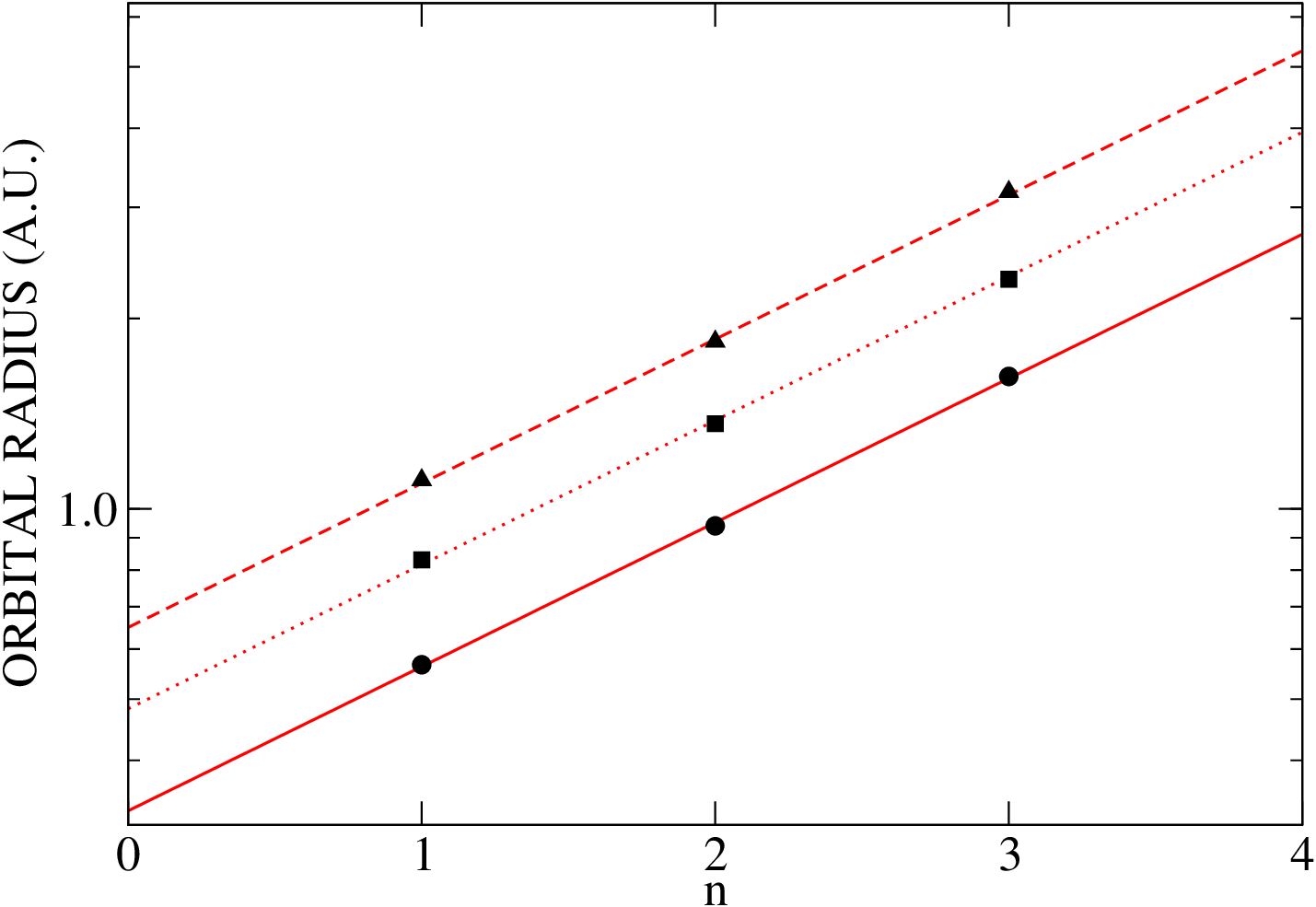}
\caption{The mean radial distance from the star of three planets for different initial conditions (symbols), after the self-organized periodic state is established. In all cases the final planetary distribution follows the Titius-Bode law $r_n \sim 1.69^n$, shown by parallel lines.
} \label{fig2c}
\vspace{0.7cm}
\end{center}
\end{figure}

We see that the dynamics is very complex, but after $t=10^4$ years, this planetary system relaxes to a 
periodic state with synodic period $T=1.6$ years, in which adjacent planetary orbits exhibit a perfect synchronization, with the time between two consecutive perihelions --- anomalistic period --- increasing in the ratio of 2:1, 
see Fig. \ref{fig2b}. The mean orbital distance of each planet in this synchronized state follows the Titius-Bode law $r_n \sim 1.69^n$, see Fig. (\ref{fig2c}).

\begin{figure}
\begin{center}
\includegraphics[scale=0.325]{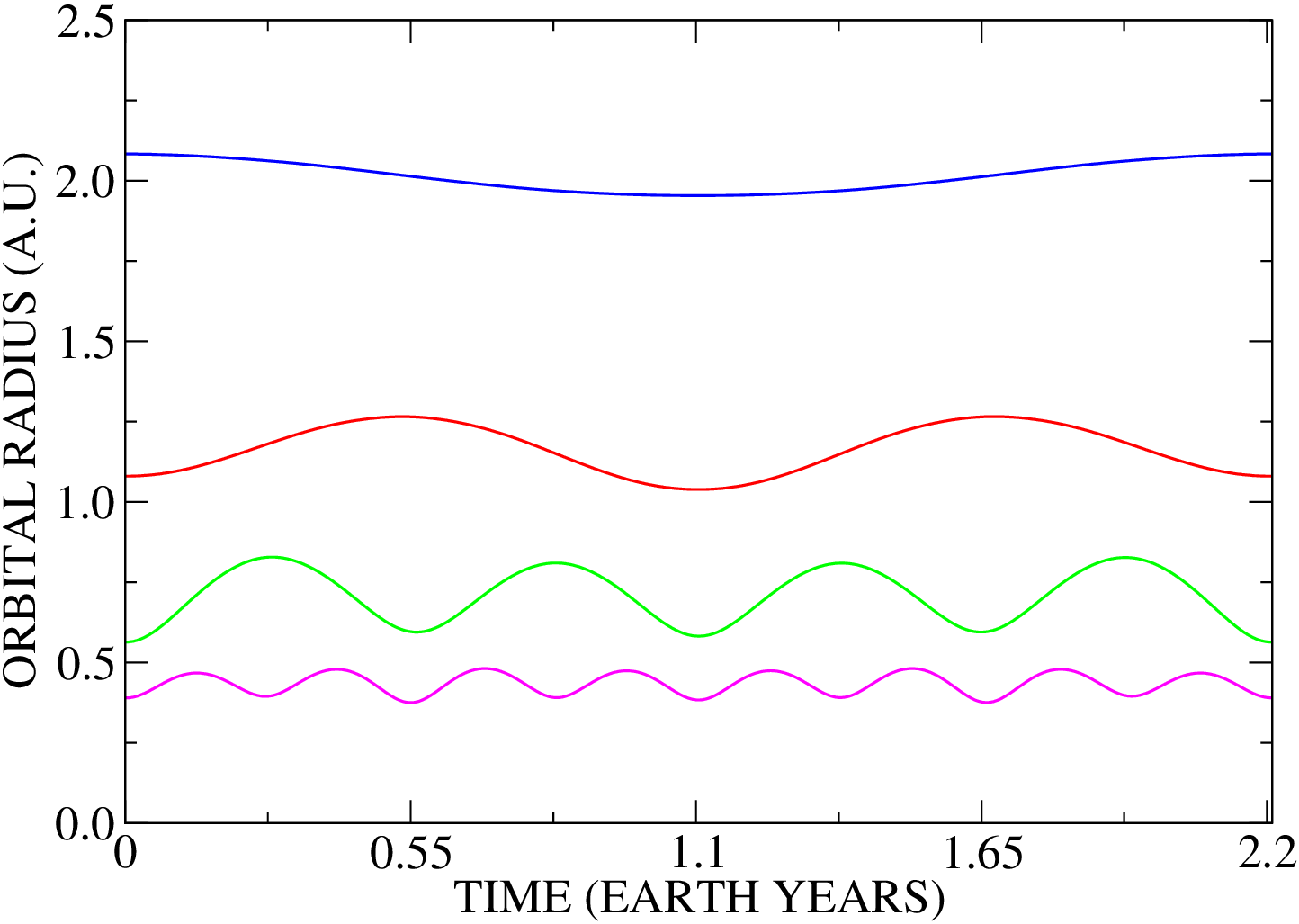}
\caption{Temporal evolution of radial coordinates of planets in a self-organized 
periodic planetary system with $4$ planets.  
Note that the orbits of adjacent planets exhibit a perfect 2:1 synchronization.
} \label{fig3}
\vspace{0.7cm}
\end{center}
\end{figure}

\begin{figure}
\begin{center}
\includegraphics[scale=0.325]{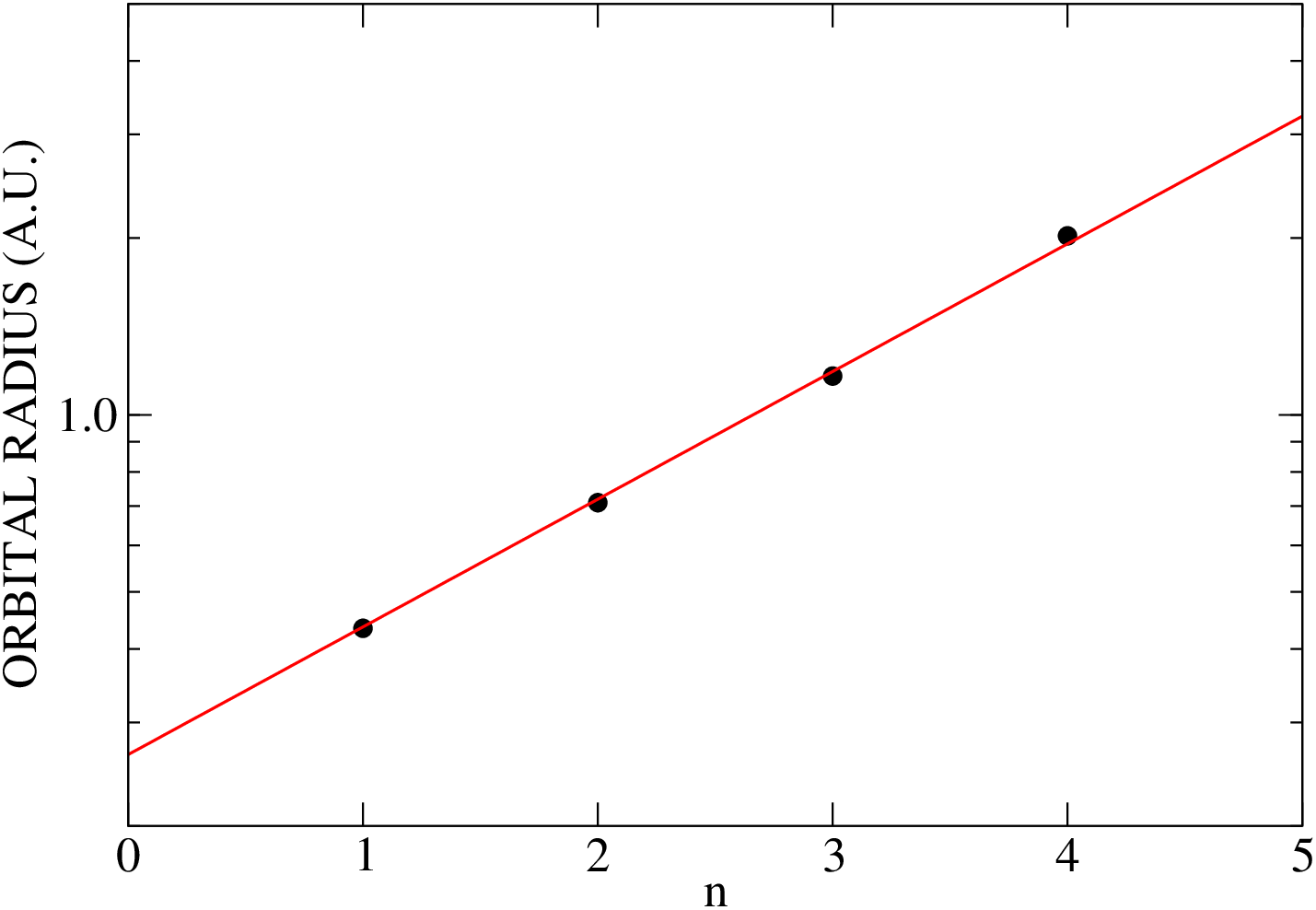}
\caption{The mean radial distance of four planets from the star, which agrees precisely with the Titius-Bode law, the solid line.} \label{fig4}
\end{center}
\end{figure}

In Fig. \ref{fig3} we show a periodic orbit to which a planetary system with $4$ planets evolves.
From the figure we see that the orbits of planets once again exhibit a perfect synchronization, with the anomalistic period of adjacent planets increasing in the ratio of 2:1. To make this observation more quantitative we define a radial and an angular winding numbers: 
\begin{eqnarray}
\label{e5}
\omega^r_i=\lim_{t \rightarrow \infty} \frac{N_{peri}(t)}{t} \\ \nonumber 
\omega^\theta_i=\lim_{t \rightarrow \infty} \frac{N_{\theta}(t)}{t},
\end{eqnarray}
where $N_{peri}(t)$ is the number of times that the orbit of planet $i$ passes through a perihelion during a time interval $t$, and $N_{\theta}(t)$ is the number of times that the 
planet $i$
completes a full rotation around the star.  It is important to keep in mind that because of the interplanetary  gravitational attraction the dynamics is very complicated, with the orbit of a planet precessing around the star. The planetary year, therefore, will not be equal to the 
anomalistic period, amount of time between two consecutive perihelions.  This dichotomy is clearly demonstrated by the normalized radial and angular winding numbers (winding number of planet $i$ divide by the winding number of the outermost planet), see Table \ref{tab1},
\begin{table}
\caption{Normalized radial and angular winding numbers of a 4 planet system.}\label{tab1}
\centering
\begin{tabular}{|c|c|c|c|c|} \hline
Planet& 1 & 2& 3& 4  \\
\hline \hline \hline
$\omega_r$ & 1.000& 2.000 & 4.000& 8.000\\ \hline
$\omega_\theta$  & 1.000 & 2.275 & 4.827&  9.929 \\ \hline
\end{tabular}
\vspace{.9cm}
\end{table}
which shows a perfect 2:1 synchronization of anomalistic periods of adjacent orbits, but no synchronization of rotational periods.
Furthermore, Fig. \ref{fig4} shows that the mean planetary distances from the star in this self-organized planetary system follow a geometric progression -- Titius-Bode law -- $r_n \sim 1.67^n$, very similar to the one observed in our solar system.  The synodic period  for this planetary system is $T=2.2$ years.

\begin{figure}
\begin{center}
\includegraphics[scale=0.325]{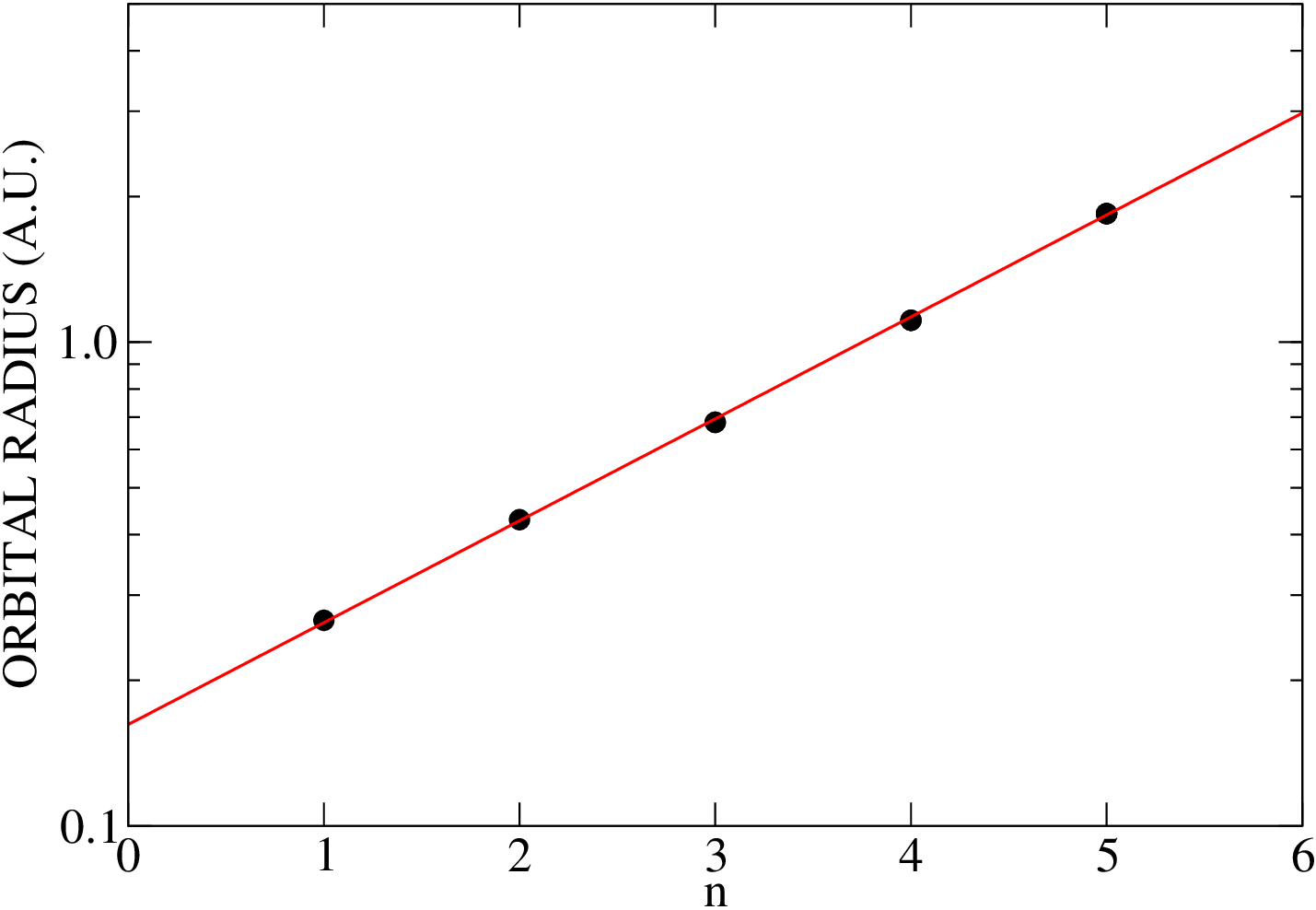}
\caption{Mean radial planetary distance from the star in a self-organized system which originally contained $9$ planets. Various planets have collided forming new planets of mass $2m$, so that only 5 distinct
radial positions appear in the plot.  These follow  the Titius-Bode distribution.
} \label{fig5}
\end{center}
\end{figure}

We next study a planetary system which starts with $9$ planets of mass $m=0.002M$ uniformly distributed from $1$ A.U. to $9$ A.U.  
After a period of dynamical evolution this system once again
self-organized into a complex periodic structure. During the evolution, planets 1 and 2 collided 
producing a new planet of mass $2m$. This planet, in turn, formed a binary with the planet 3.   Planets 4 and 5 also collided forming a new planet of mass $2m$, as did planets 6 and 7.  The resulting planetary system contains 6 planets with synodic period $T=6.5$ years. In spite of a distinct mass distribution and a very complex dynamics,  the planetary distances are found to, once again, follow a geometric progression $r_n \sim 1.63^n$, see Fig. \ref{fig5}.  Furthermore, the radial winding numbers indicate a perfect 2:1 syncronization between anomalistic periods, while the orbital periods do not show any clear structure, see Table \ref{tab2}. Finally, we should note that it is very difficult to obtain large planetary systems, even if we start with a very large number of planetesimals, very fast most of them will either fall into the sun, coalesce, or will be ejected.   The final planetary system will have only a small number of planets.  In spite of a diligent effort we were not able to find a stable planetary system with more than 6 planets.

\begin{table}
\caption{Radial and angular winding numbers of a self-organized system with 9 planets.}\label{tab2}
\centering
\begin{tabular}{|c|c|c|c|c|c|c|c|c|c|} \hline
Planet& 1 & 2& 3& 4& 5& 6& 7& 8& 9 \\
\hline \hline \hline
$\omega_r$ & 1.000& 1.000 & 1.000 & 2.000& 2.000&  4.000& 4.000& 8.000& 16.000\\ \hline
$\omega_\theta$  & 1.000& 1.000 & 1.000 & 2.146 & 2.146&  4.439& 4.439&  9.024& 18.195\\ \hline
\end{tabular}
\end{table}

\section{Conclusions}

We have shown that stability of planetary systems is intimately connected with the orbital arrangement of planets.
An arbitrary initial distribution of planets is susceptible to catastrophic events in which
planets  are ejected from the planetary system or collide with each other. These catastrophic events are 
an unavoidable consequence of chaotic dynamics and of Arnold diffusion characteristic of celestial mechanics in the $t \rightarrow \infty$ limit. We note that even if the planets are placed at radial positions consistent with the Titius-Bode law, a planetary system will still, in general, be unstable unless the orbits of planets are properly synchronized and the dynamics is periodic.  In this paper we presented a mechanism which leads to self-organization of a planetary system into a stable periodic state.  The mechanism proposed is probably not unique and should rather be viewed as a proof of concept which demonstrates that energy non-conserving perturbations can drive a planetary system into a self-organized periodic state 
from an {\it arbitrary} initial condition.   In such state  anomalistic periods between all planets are synchronized,  while orbital periods do not indicate any  synchronous structure. When the anomalistic periods between the radially adjacent planets are synchronized  in 2:1 ratio, the mean orbital
distance is found to follow a geometric progression, $r_n \sim 1.7^n$, the same as the one observed in our solar system.  In principle, however, there is no {\it a priori} reason why all the planets should have 2:1 synchronization, and indeed other synchronized states are possible. For such planetary systems Titius-Bode law will not be valid. We stress again that it is the anomalistic, and not the orbital periods, which show synchronous behavior in the self organized state.  Indeed, if orbital periods would be synchronized, the explanation of the Titus-Bode law would be quite straight forward.  For small planetary masses, the planetary year is related to the length of the semi-major axis through the Kepler's law $T^2 \sim a^3$, if the planetary years would be locked in 2:1 resonance, the ratio of semi-major axis would then follow a geometric progression $r_n \sim 2^{2 n/3} \approx 1.5874^n$, which is very similar to the observed Titus-Bode law.
However, the Tables 1 and 2, show that there is no synchrony of orbital periods and only anomalistic periods that are synchronized.  Therefore, this simple argument can not be used to account for the orbital structure inside the self-organized state observed in our simulations.

\section{Acknowledgments}

This work was partially supported by the CNPq, National Institute of Science and Technology Complex Fluids INCT-FCx, and by the US-AFOSR under the grant 
FA9550-16-1-0280.

\end{document}